# Experimental observation of non-Hermitian higher-order skin interface states in topological electric circuits

Bin Liu[1], Yang Li[1], Bin Yang[1], Xiaopeng Shen[1], Yuting Yang[1, *], Zhi Hong Hang[2, 4] and Motohiko Ezawa[3, *]

[1]*School of Materials Science and Physics, China University of Mining and Technology, Xuzhou 221116, China*

[2]*School of Physical Science and Technology & Collaborative Innovation Center of Suzhou Nano Science and Technology, Soochow University, Suzhou 215006, China*

[3]*Department of Applied Physics, University of Tokyo, Hongo 7-3-1, 113-8656, Japan*

[4]*Institute for Advanced Study, Soochow University, Suzhou 215006, China*

*Corresponding authors (Yuting Yang, email: yangyt@cumt.edu.cn; Motohiko Ezawa, ezawa@ap.t.u-tokyo.ac.jp)

Abstract: The study of topological states has developed rapidly in electric circuits, which permits flexible fabrications of non-Hermitian systems by introducing non-Hermitian terms. Here, nonreciprocal coupling terms are realized by utilizing a voltage follower module in non-Hermitian topological electric circuits. We report the experimental realization of one- and two- dimensional non-Hermitian skin interface states in electric circuits, where interface states induced by non-Hermitian skin effects are localized at the interface of different domains carrying different winding numbers. Our electric circuit system provides a readily accessible platform to explore non-Hermitian-induced topological phases, and paves a new road for device applications.

## I. Introduction

Topological insulators [1,2] and higher-order topological insulators [3-9] have attracted intense research interest. They are characterized by the emergence of topologically protected edge states which are immune to perturbations. They are also realized in photonic [10-14], acoustic [15-17] and electric circuit systems [18-22]. A real physical system is often open and has non-negligible interactions with the environment and hence the Hamiltonian is non-Hermitian, where eigenvalues are complex in general. The interplay between non-Hermitian and topological phases can induce many novel phenomena [23-29]. One of recent discovers is the breakdown of conventional bulk-boundary correspondence, which is derived from non-Hermitian skin effects [30-33], where all eigenstates of an open system are localized near the edge. The Hatano-Nelson model is a nonreciprocal hopping model [34], which hosts non-Hermitian skin states. A non-Bloch bulk-boundary correspondence has been discovered to describe topological properties of one-dimensional (1D) non-Hermitian system

[30,35-37]. Non-Hermitian skin effects are generalized to higher-order non-Hermitian skin effects [38-43]. Non-Hermitian systems have been investigated in photonic crystals [44,45], acoustic crystals [46,47] and quantum walks [48]. Especially, non-Hermitian skin effects and higher-order skin effects experimentally observed in acoustic system [49,50]. However, a typical nonreciprocal system, the Hatano-Nelson model, has not yet been observed in 2D electric circuit experiments. In addition, the comparison between the open-chain circuit and the closed-loop circuit has not been explored experimentally. Furthermore, there is no experimental observation of skin interface states theoretically proposed in Ref. [39].

Electric circuits have been the subject of recent theoretical and experimental interest because of the convenience and flexibility with which they can be designed and fabricated to realize different topological phases [18-22,51], such as corner modes [20,22,52], nonlinear topological boundary states [53,54] and four-dimensional topological insulators [55-57]. One can conveniently introduce a non-Hermitian term in an electric circuit [58-63], for example, by adding an operational amplifier (Op-Amp) to form a negative-impendence converter with current inversion as a gain element, and a positive resistor as a loss element in the Su-Schrieffer-Heeger model [58-61]. On the other hand, nonreciprocal hoppings can be realized by using Op-Amps with unidirectional current flow [62-64]. Non-Hermitian skin effects have been observed in electric circuits [60,63-66].

In this work, we experimentally demonstrate higher-order skin interface states in a non-Hermitian topological circuit by utilizing the Hatano-Nelson model based on the theoretical proposal [39]. The nonreciprocal coupling terms are realized by utilizing a voltage follower module [60], and adjusted by capacitors. The circuit Laplacian directly corresponds to the Hamiltonian of the physical system [18-22], where the voltages are obtained by simulations and experiments. 1D, 2D and high-order Non-Hermitian topological circuits are realized through steady-state and time-domain simulations. Experiments are carried out to verify the existence of interface state and corner mode in non-Hermitian circuits. Our experimental results are in good agreement with simulations in frequency spectra and voltage distributions. Our designs might pave a road to new designs of non-Hermitian topological circuits.

## II. 1D non-Hermitian circuit

We start with an interface made of the 1D Hatano-Nelson model as shown in Fig. 1(a), which is a most typical nonreciprocal system. The asymmetric hopping amplitudes between adjacent sites are indicated by rightward coupling $t_r$ and leftward coupling $t_l$. The topological phase of a non-Hermitian system is characterized by the winding number $w = \sum_{n=1}^{N} \int_{-\pi}^{\pi} \frac{dk}{2\pi} \partial_k \arg E_n(k)$ (N is the total number of bands and $\arg E_n(k)$ is the argument of the complex energy $E_n(k)$) [27], where $w=1$ for the region I ($|t_r|<|t_l|$) and $w=-1$ for the region II ($|t_r|>|t_l|$). All bulk eigenstates are localized at the interface of the two regions with inverted winding numbers, and decay exponentially. The winding directions are indicated by arrows in Fig.1(a). This model can be realized in an electric circuit as shown in Fig. 1(d). Each lattice site is composed of a LC resonant tank with inductance $L_0$ and capacitance $C_0$. The nonreciprocal coupling amplitude is realized by the capacitance $C_g$ with a voltage follower, where current flows unidirectionally. The voltage follower is combined with an Op-Amp, resistors $R_a$, $R_b$ and capacitance $C_a$.

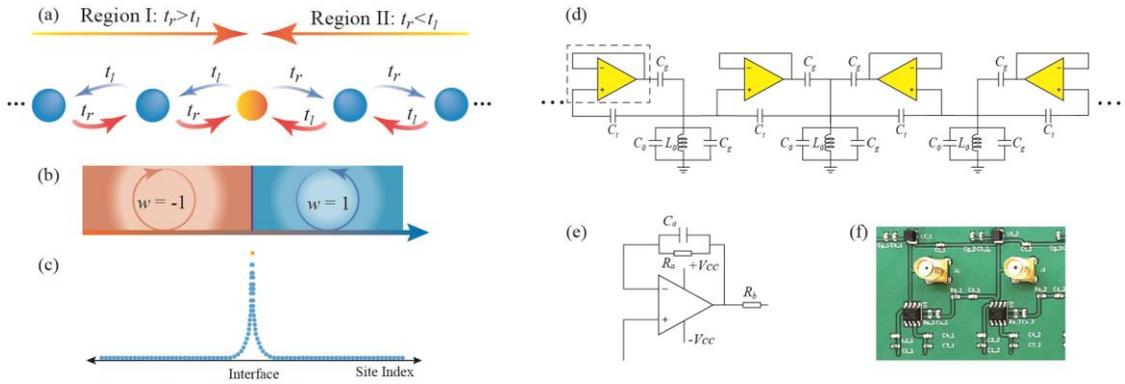

FIG. 1 (a) Schematic diagram of a nonreciprocal non-Hermitian model with two topological distinct regions. (b) The winding number $w=1$ for the region I ($|t_r|>|t_l|$), and $w=-1$ for the region II ($|t_r|<|t_l|$). (c) The eigenstate exponentially decays from the interface of the two regions. (d)

Schematic diagram of a nonreciprocal non-Hermitian circuit. (e) Schematic of a voltage follower [60]. (f) Photograph of the printed circuit board of two unit cells in the non-Hermitian circuit.

Kirchhoff's laws are applied to analyze properties of the circuit (see the detailed derivation in Appendix A). Two edges of a 1D open chain circuit with one region are connected to the ground, and lattice sites are indexed by an integer $n$. A mode with angular frequency $\omega$ for an open chain circuit satisfies

$$(H - 2\frac{C_t + C_g}{C_0}) \begin{pmatrix} v_1 \\ v_2 \\ v_3 \\ v_4 \\ \vdots \end{pmatrix} = (1 - \frac{\omega_0^2}{\omega^2}) \begin{pmatrix} v_1 \\ v_2 \\ v_3 \\ v_4 \\ \vdots \end{pmatrix}, \tag{1}$$

and the matrix $H$ denotes the Hamiltonian of the circuit structure,

$$H = \begin{pmatrix} 0 & t_l & & & \\ t_r & 0 & t_l & & \\ & t_r & 0 & \ddots & \\ & & \ddots & \ddots & \end{pmatrix}, \tag{2}$$

where $v_n$ is the complex voltage on the lattice site $n$, $\omega_0 = 1/\sqrt{L_0 C_0}$ is the characteristic frequency of the circuit, $t_r = (C_t + C_g)/C_0$ and $t_l = C_t/C_0$ are nonreciprocal hopping terms. We also consider a closed loop circuit, in which the left side of the region I connects with the right side of the region II, where it has two interfaces. The Hamiltonian for the closed loop circuit is

$$H = \begin{pmatrix} 0 & t_{l1} & 0 & \cdots & \cdots & \cdots & \cdots & 0 & t_{r2} \\ t_{r1} & 0 & t_{l1} & 0 & \cdots & \cdots & \cdots & \cdots & 0 \\ \ddots & \ddots & \ddots & \ddots & \ddots & \ddots & \ddots & \ddots & \ddots \\ 0 & \cdots & t_{r1} & 0 & t_{l1} & 0 & \cdots & \cdots & 0 \\ 0 & \cdots & 0 & t_{r2} & 0 & t_{l2} & 0 & \cdots & 0 \\ 0 & \cdots & 0 & 0 & t_{r2} & 0 & t_{l2} & \cdots & 0 \\ \ddots & \ddots & \ddots & \ddots & \ddots & \ddots & \ddots & \ddots & \ddots \\ 0 & \cdots & \cdots & \cdots & \cdots & 0 & t_{r2} & 0 & t_{l2} \\ t_{l2} & 0 & \cdots & \cdots & \cdots & \cdots & 0 & t_{r2} & 0 \end{pmatrix}. \tag{3}$$

By solving the eigenvalue of the Hamiltonian matrix, we obtain the frequency spectra of the designed circuit. The eigenfrequency is shown as a function of $t_r/t_l$ in Figs. 2(a) and 2(b), corresponding to the open chain circuit and the closed loop circuit, respectively. The hopping amplitude $t_l$ is fixed when we choose $C_0 = 470\ pF$ and $C_t = 680\ pF$, while $t_r$ varies with $C_g$ accordingly. All intersection of a red line corresponds to edge (interface) states in the circuit, induced by the non-Hermitian skin effect. The case with $C_g = 470\ pF$ corresponding to $t_r = 2.45$ and $t_l = 1.45$ is shown in Figs. 2(c), (d) and (e).

The energy spectra with the periodic boundary condition (PBC) is:

$$E = \frac{2C_t + C_g}{C_0 + 2C_g + 2C_t}\cos k + i\frac{-C_g}{C_0 + 2C_g + 2C_t}\sin k, \qquad (4)$$

where $E = \frac{\omega_{es}^2}{\omega^2} - 1$ and $\omega_{es} = \sqrt{L_0(C_0 + 2C_g + 2C_t)}$ (see the detailed derivation in Appendix A). As shown in Fig. 2(c), the energy spectrum is elliptical in the complex plane in the case that $t_r = 2.45$ and $t_l = 1.45$. When the ratio $t_r/t_l$ in the region I and $t_l/t_r$ in the region II are identical, two elliptical energy spectra are degenerated and encircle along inverted directions, which give rise to the winding number $w = -1(+1)$. The energy spectra in the open boundary condition (OBC) is shown in Figs. 2(d) and 2(e), which collapse to a line with the zero imaginary part.

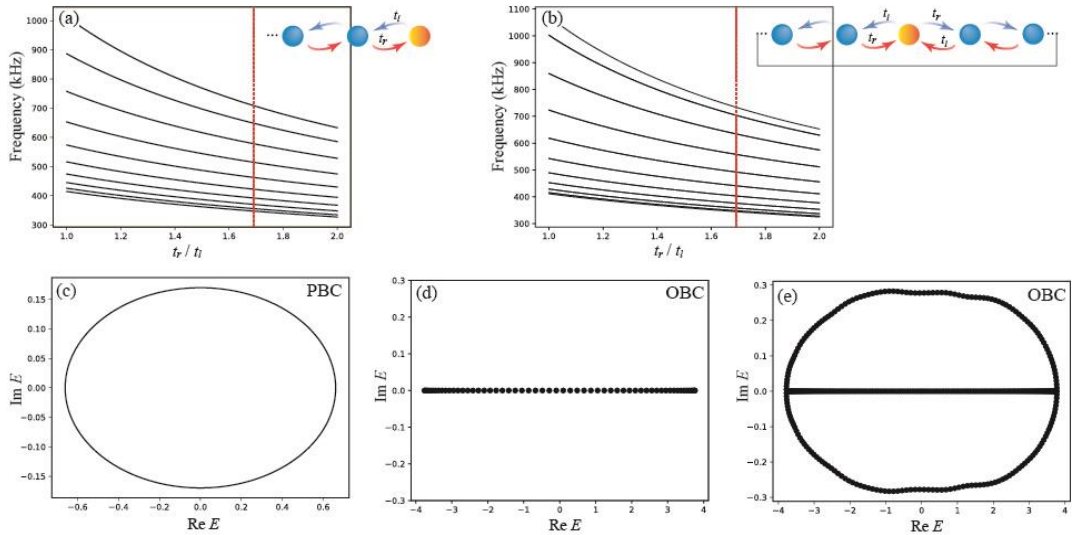

FIG. 2 (a) and (b) Frequency spectra of the nonreciprocal non-Hermitian in the open-chain circuit and the closed-loop circuit, respectively. Parameters are set as $C_0 = 470\ pF$, $L_0 = 47\ \mu H$, and $C_t = 680\ pF$. The hopping ratio is $t_r/t_l = 1 + C_g/C_t$. (c) Elliptical energy spectra with the PBC. (d) and (e) Complex energy spectra of the open-chain circuit and the closed-loop circuit with the OBC respectively, which corresponds to red dashed lines in (a) and (b) with $C_g = 470\ pF$ i.e. $t_r = 2.45$ and $t_l = 1.45$.

The designed 1D nonreciprocal circuit reveals a noteworthy non-Hermitian topological feature. This feature is characterized by a distinctive skin effect, which is localized on the edge or interface of a circuit. To observe the skin effect, we implement a closed loop circuit using a FR4 printed circuit board (PCB) as pictured in Fig. 3(a). The PCB contains additional traces used to connect two boundaries. The created closed loop sample possesses 20 sites and two distinct interfaces i.e. nodes 11 and 20. The circuit parameters are chosen to be the same as those in Fig. 2. The voltage amplitude of an interface denoted by a red arrow in Fig. 3(a) at different frequencies is simulated by LTspice software and displayed in Fig. 3(b). There are some resonance peaks, which agree well with intersection points of frequencies in Fig. 2(b). In simulations, the Equivalent Series Resistance (ESR) of capacitors, inductors and resistors are taken into consideration by using the same values as the real devices used in our experiment. We include the 5% randomness of circuit components, whose effects are simulated by the Monte Carlo method. The results are shown by ten curves in Fig. 3(b). The ideal circuit simulations without resistors and randomness are provided in Appendix B, in which the amplitude peaks are very high and sharp.

In experimental measurements, the input signal is produced by a function generator (Tektronix AFG31000) and the output signal is obtained by an oscilloscope (Tektronix TBS2000B) via the preset SMA ports on each site in a circuit. A DC power supplier is used to provide 15V DC voltage for Op-Amps (LM6171) in the electric circuit. In order to effectively excite interface states, we place a source with 1V sine wave at site 10 (marked by a red star) close to an interface. The experimental results of voltage amplitudes are shown in Fig. 3(d), in which the peaks appear in the same high frequency range with simulations to display interface states. There is a difference for the first peak in Fig. 3(b) and 3(d), which is sensitive to the randomness variation of circuit parameters, and the frequency shift of other peaks are accounted for non-ideal Op-Amps in the simulations and decoupling capacitors in real circuits. The amplitude peaks in measurements are not very sharp, which are caused by background resistance of the fabricated PCB and component resistance. We investigate voltage amplitudes at different sites in experimental and simulated results depicted in the Figs. 3(c) and 3(e). Even though there are two interfaces in the closed loop circuit, the strongest voltage is only localized at the middle interface. Our experimental results agree well with simulations, and directly demonstrates the phenomenon of skin effects in the non-Hermitian circuit.

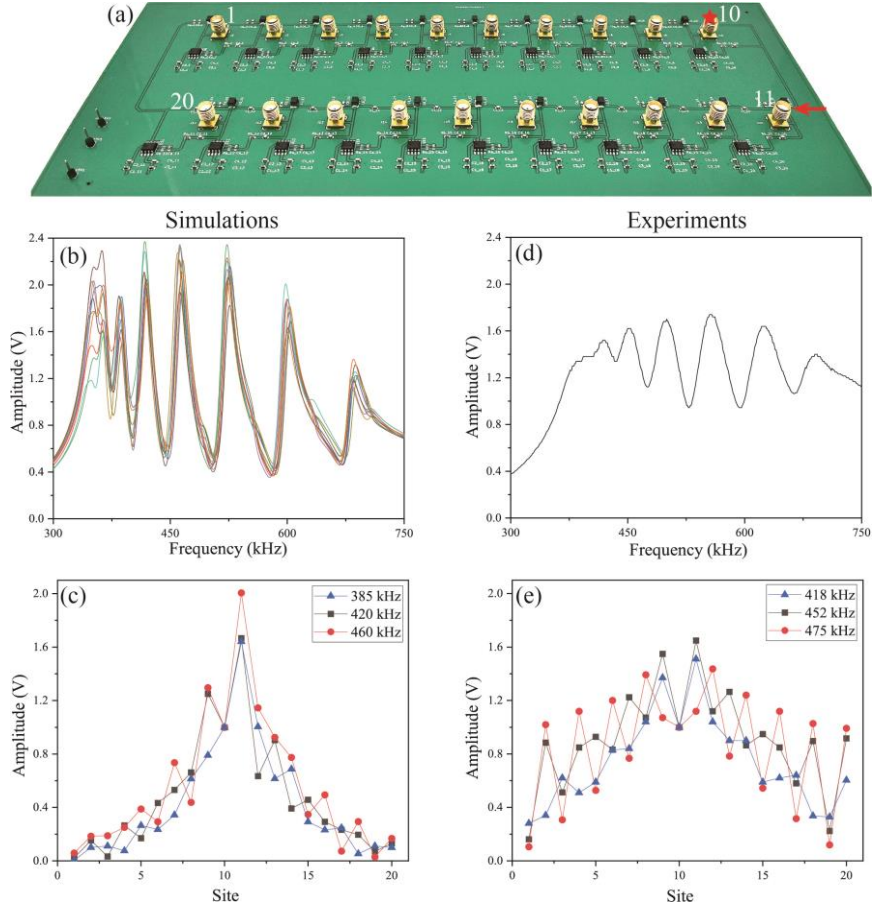

FIG. 3 (a) Photograph of a fabricated closed loop circuit containing 20 nodes with $C_0$ = 470 pF, $C_g$ = 470 pF, $L_0$ = 47 μH, $C_t$ = 680 pF. The circuit has two interfaces at nodes 11 and 20. (b) and (d) Voltage amplitude at different frequencies for node 11 in experimental measurements and simulations. Different curves in (b) presents 10-times simulated results under 5% randomness in circuit components. (c) and (e) Voltage distribution at different sites in experimental and simulated results, showing strong interface localization at 418, 452 and 475 kHz in experiments and 385, 420 and 460 kHz in simulations. The location of an excited source is placed at node 10 both in simulations and experiments.

## III. 2D non-Hermitian circuit

We generalize the configuration of the 1D Hatano-Nelson model to a 2D nonreciprocal circuit with alternating coupling terms $t_r$ and $t_l$ in the $x$ and $y$ directions, as schematically illustrated in Fig. 4(a). The circuit structure composed of 9×9 sites includes two domain walls whose winding numbers are $(w_x, w_y) = (\mp 1, \pm 1)$, and four boundaries are connected to ground. Figure 4(b) shows voltage amplitudes at different frequencies in simulations, which are selected at three sites on the interface. The voltage resonance peaks correspond to the localized interface states. We performed time-domain simulations of a circuit structure by using a Gaussian input signal $S = e^{-\frac{(t-t_0)^2}{dt^2}} \cos \omega t$ with time delay $t_0 = 25$ $\mu s$, width $dt = 25$ $\mu s$ and frequency $f = 309$ kHz, as shown in the inset in Fig. 4(c). The Gaussian pulse impinges on the circuit from the upper-rightmost corner site. Excited interface states travel through the interface and reach the bottom corner around 65 $\mu s$. The output signal remains the wave shape unchanged, which indicates the loss is very low in the propagation of interface states.

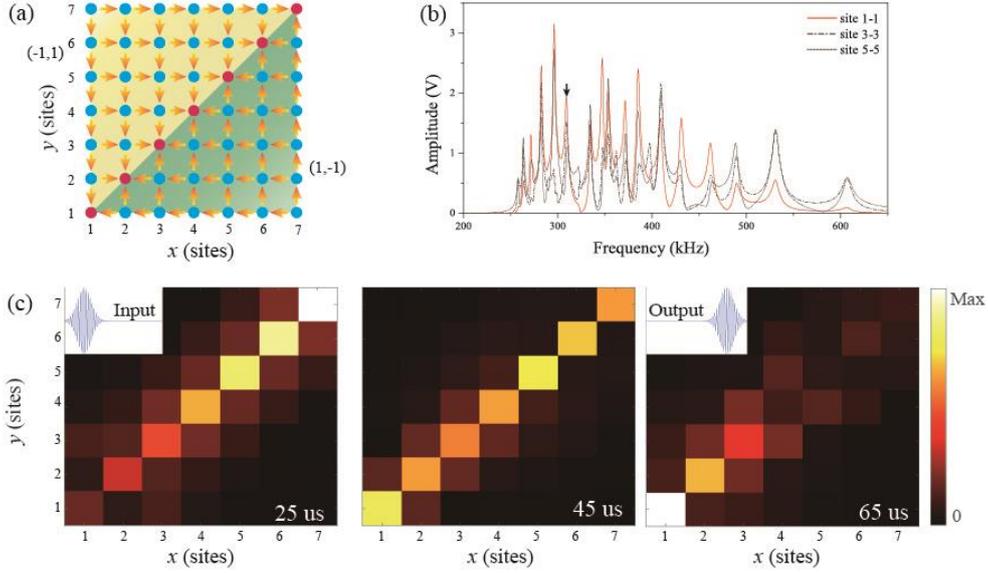

FIG 4 (a) Schematic of the 2D nonreciprocal non-Hermitian circuit with diagonal interface state. (b) Voltage amplitude as a function of frequencies at three sites of the interface in simulations. (c) Simulated propagation of the interface state along the diagonal interface nodes excited by a Gaussian pulse at 309 kHz marked by a yellow arrow in (b). The excited source is located at upper-rightmost site, and the circuit parameters are set to the same value as in the 1D non-Hermitian circuit.

We then experimentally demonstrate the skin interface state in the 2D nonreciprocal non-Hermitian system. The experimental sample of the electric circuit with 7×7 sites is shown in Fig. 5(a). The location of excited source is denoted by a star at site (7, 7). Voltage amplitudes at all sites are experimentally measured and have a function with different frequencies. The amplitude peaks around 400 kHz at the interface sites are displayed in Fig. 5(b). The steady state distribution of the voltage amplitude is displayed in Fig. 5(c) at 395 kHz corresponding to the frequency peak in Fig. 4(d). Strong amplitudes localize at the interface (marked by a white-dashed line in Fig.5(c) and exponentially decrease in the $x$ and $y$ directions, which indicates the skin effect in the 2D nonreciprocal circuit. The phenomenon of skin effects more than three dimensions will be realized in a higher dimensional system based on the electric circuit structure.

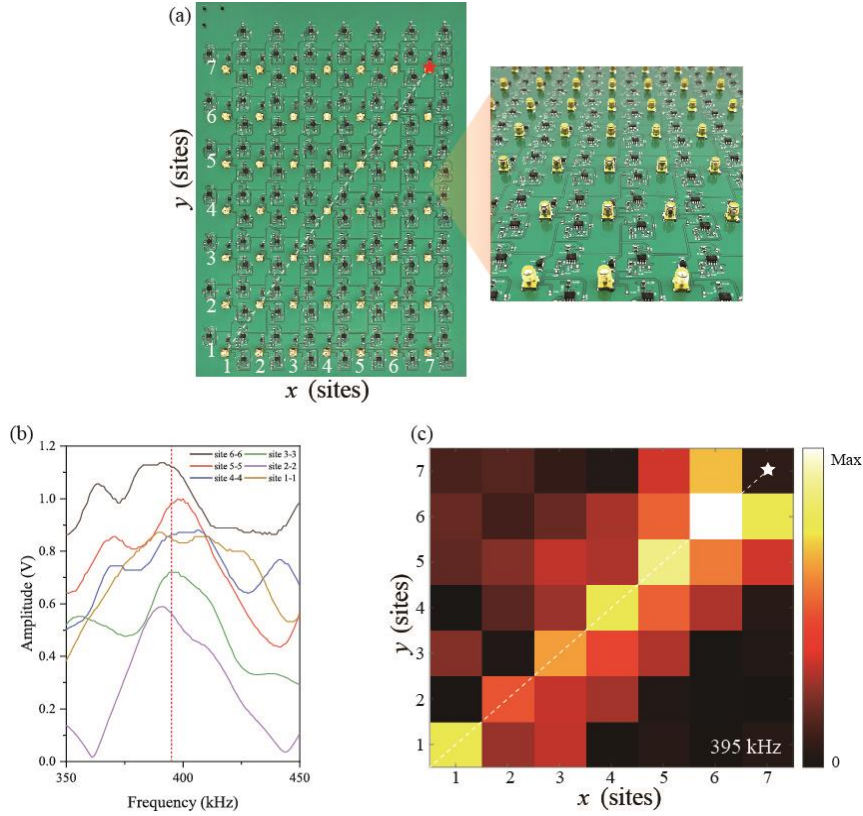

FIG. 5 (a) Fabricated sample of a 7x7 circuit with $C_0 = 470pF$, $C_g = 470pF$, $L_0 = 47\mu H$, $C_t = 680pF$. A white line indicates the diagonal interface, and a red star denote the location of excited source. (b) Voltage amplitude at different frequencies of a 2D nonreciprocal non-Hermitian circuit. Each curve represents the voltage variation of a site on the diagonal of the circuit. (c) Experimental voltage distribution in the steady state in 2D nonreciprocal non-Hermitian circuit at 395 kHz indicated by a red line in (b).

Let us construct a non-Hermitian electric circuit with four domains, which possesses four different topological phases. The skin corner mode emerges in the central part. This is the second-order non-Hermitian skin interface state. Figure 6(a) shows the schematic and experimental sample of the electric circuit possessing 9×9 sites with the winding number of four domain walls in the $x$ and $y$ directions. The voltage amplitudes at corner and bulk sites is as a function of frequencies in experimental measurements as displayed in Fig. 6(b). The experimental result at 308 kHz corresponding to the dashed line in Fig. 6(b) displays the strong voltage amplitude concentrates in the central part in Fig. 6(c), which demonstrates the existence of the skin corner mode in the non-

Hermitian circuit. To excite the corner mode effectively, the excited source is placed near the corner site, indicated by a white star.

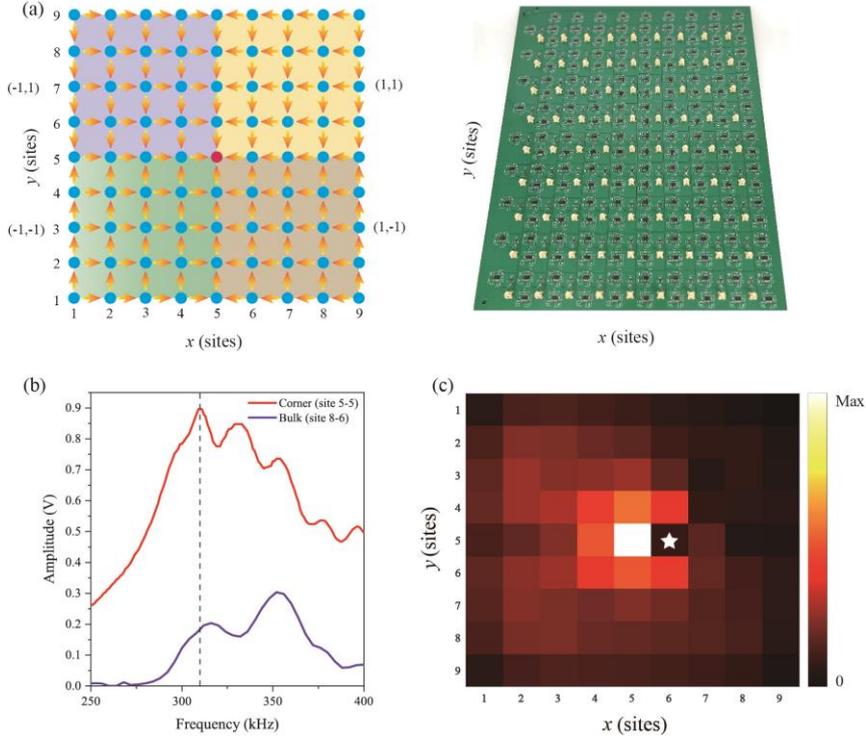

FIG. 6 (a) Schematic and fabricated sample of the corner mode realized in the 2D nonreciprocal non-Hermitian circuit. The circuit parameters are $C_0 = 470pF$, $C_g = 470pF$, $L_0 = 47\mu H$, $C_t = 680pF$. (b) Voltage amplitude at different frequencies of a 2D nonreciprocal non-Hermitian circuit in experiments. The red and blue curves represent the corner and bulk site of the circuit, respectively. (c) Experimental measurements of the voltage amplitude distribution of the corner mode in 2D non-Hermitian circuit at 308 kHz corresponding to the dashed line in (b).

## IV. Conclusion

With the help of Op-Amp components, nonreciprocal hopping terms are materialized in non-Hermitian circuits. Based on the Kirchhoff law possessing one-to-one correspondence with the tight-binding Hamiltonian, we have theoretically derived the real and imaginary parts of the energy spectrum of the circuit under PBC and OBC. We have simulated the time-domain propagation of a pulse signal of non-Hermitian skin interface states. Furthermore, we have experimentally demonstrated skin interface states in 1D and 2D non-Hermitian topological circuits. The influencing factors of experimental results in electric circuits are also analyzed. For example, the randomness and resistance give rise to a frequency shift and a small $Q$ factor, which provide a reference for future circuit experiments. Topological circuits provide convenient platforms with a good vision to experimentally demonstrate topological physics. Our work presents fascinating phenomena of non-Hermitian skin effects in different dimensions of topological circuits, which indicates that topological circuit is an effective platform for future Non-Hermitian skin effect is very sensitive to boundary perturbations, so that many non-Hermitian topological devices, for example, the sensor is theoretically proposed [67,68] and experimentally realized in electric circuits [69], which has a high-level sensitivity and a strong robustness. Therefore, the in-depth study of non-Hermitian circuit provides a new idea to realize potential device applications by combing non-Hermitian physics and integrated circuits.

## Acknowledgments

*This work is supported by the National Natural Science Foundation of China (Nos. 12004425 and 12274315), the Natural Science Foundation of Jiangsu Province (No. BK20200630), a project funded by the Priority Academic Program Development of Jiangsu Higher Education Institutions (PAPD), the Basic Research Program of Xuzhou (KC22016), the Key Academic Discipline Project of China University of Mining and Technology (No. 2022WLXK06), CREST, JST (Grants No. JP-MJCR20T2) and the Grants-in-Aid for Scientific Research from MEXT KAKENHI (Grant No. 23H00171).*

## APPENDIX A: CIRCUIT EQUATIONS

By applying the Kirchhoff's laws to the node $n$, the circuit equation of a non-Hermitian open-chain electric circuit is obtained as

$$I_n = V_n \cdot i\omega(-\frac{1}{\omega^2 L_0} + C_0 + C_g) \tag{A1}$$

$$I_n + (V_n - V_{n+1}) \cdot i\omega C_t - (V_{n-1} - V_n) \cdot i\omega(C_t + C_g) = 0. \tag{A2}$$

Combining Eq. (A1) and Eq. (A2) to eliminate $I_n$, we obtain the following equation,

$$V_n \cdot i\omega(-\frac{1}{\omega^2 L_0} + C_0 + 2C_g + 2C_t) = V_{n-1} \cdot i\omega(C_t + C_g) + V_{n+1} \cdot i\omega C_t \tag{A3}$$

As we set $\omega_0 = 1/\sqrt{L_0 C_0}$, $t_r = \frac{C_t + C_g}{C_0}$, $t_l = \frac{C_t}{C_0}$, we obtain

$$V_n \cdot (1 - \frac{\omega_0^2}{\omega^2}) = V_{n-1} \cdot t_r - V_n \cdot 2t_r + V_{n+1} \cdot t_l. \tag{A4}$$

For an open chain circuit with one region, we take the left boundary into consideration, and we obtain

$$V_1 \cdot (1 - \frac{\omega_0^2}{\omega^2}) = -V_1 \cdot 2t_r + V_2 \cdot t_l. \tag{A5}$$

Hence, we have

$$\begin{pmatrix} 2t_r & t_l & & & \\ t_r & 2t_r & t_l & & \\ & t_r & 2t_r & t_l & \\ & & t_r & 2t_r & \ddots \\ & & & \ddots & \ddots \end{pmatrix} \begin{pmatrix} v_1 \\ v_2 \\ v_3 \\ v_4 \\ \vdots \end{pmatrix} = (1 - \frac{\omega_0^2}{\omega^2}) \begin{pmatrix} v_1 \\ v_2 \\ v_3 \\ v_4 \\ \vdots \end{pmatrix}. \tag{A6}$$

In this case, the Hamiltonian is given by

$$H = \begin{pmatrix} 0 & t_l & & & \\ t_r & 0 & t_l & & \\ & t_r & 0 & t_l & \\ & & t_r & 0 & \ddots \\ & & & \ddots & \ddots \end{pmatrix}. \tag{A7}$$

The Hamiltonian for a closed-loop circuit with two regions is

$$H = \begin{pmatrix} 0 & t_{l1} & 0 & \cdots & \cdots & \cdots & \cdots & 0 & t_{r2} \\ t_{r1} & 0 & t_{l1} & 0 & \cdots & \cdots & \cdots & \cdots & 0 \\ \ddots & \ddots & \ddots & \ddots & \ddots & \ddots & \ddots & \ddots & \ddots \\ 0 & \cdots & t_{r1} & 0 & t_{l1} & 0 & \cdots & \cdots & 0 \\ 0 & \cdots & 0 & t_{r2} & 0 & t_{l2} & 0 & \cdots & 0 \\ 0 & \cdots & 0 & 0 & t_{r2} & 0 & t_{l2} & \cdots & 0 \\ \ddots & \ddots & \ddots & \ddots & \ddots & \ddots & \ddots & \ddots & \ddots \\ 0 & \cdots & \cdots & \cdots & \cdots & 0 & t_{r2} & 0 & t_{l2} \\ t_{l2} & 0 & \cdots & \cdots & \cdots & \cdots & 0 & t_{r2} & 0 \end{pmatrix} \tag{A8}$$

The above derivation shows the Hamiltonian with the OBC, and the dispersion relation with the PBC is deduced as follows:

By combining Eq. (A4) and the Bloch theorem $V_{n+1} = V_n \cdot e^{ik}$, we obtain,

$$V_n \cdot (-\frac{1}{\omega^2 L_0} + C_0 + 2C_g + 2C_t) = V_n \cdot (C_t + C_g) e^{-ik} + V_n \cdot C_t e^{ik}. \tag{A9}$$

Setting $\omega_{es} = \sqrt{L_0(C_0 + 2C_g + 2C_t)}$, and eliminating the $V_n$, Eq. (A9) yields a result analogous to the dispersion relation:

$$\frac{\omega_{es}^2}{\omega^2} - 1 = \frac{2C_t + C_g}{C_0 + 2C_g + 2C_t} \cos k - i \frac{C_g}{C_0 + 2C_g + 2C_t} \sin k \tag{A10}$$

# APPENDIX B: SIMULATION IN 1D NON-HERMITIAN CIRCUIT

We present an ideal simulation of 1D non-Hermitian circuit structure. All components in the simulated circuit have the same parameters with that in Fig. 3, but do not have resistance and randomness. The excited source is set at the node 1. Fig. 7(a) shows some very high and sharp resonance peaks for the interface node. The localization of voltage amplitude at the interface site indicates the existence of the interface state. There is a normalization of the amplitude in Fig. 7(b) applied on different frequencies because amplitude peaks of interface node in different frequencies may vary greatly in the ideal circuit.

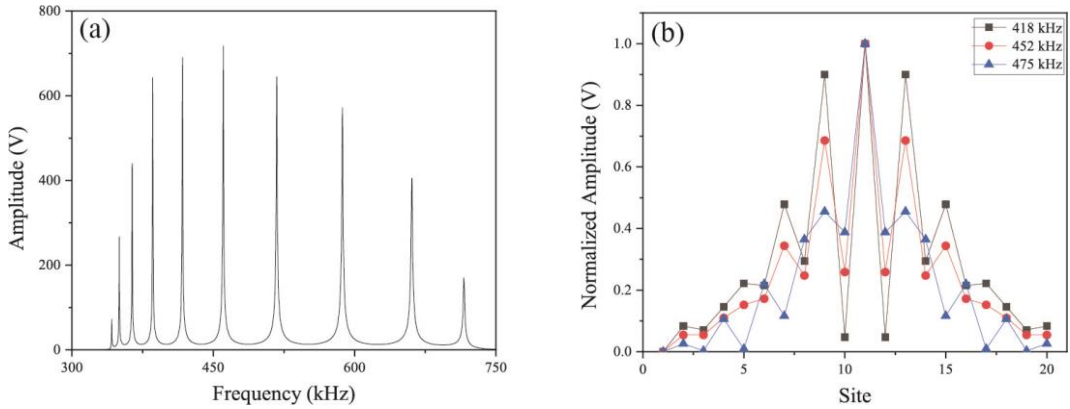

Fig. 7 (a) Voltage amplitude at different frequencies for the interface node 11 in simulations. (b) Voltage distribution at different sites in simulated results, showing strong interface localization at the same frequencies as experiments.